\documentclass{JHEP}
\usepackage{amsmath}
\usepackage{amssymb}
\usepackage{graphicx}
\usepackage{dcolumn}
\usepackage{bm}

\def\spose#1{\hbox to 0pt{#1\hss}}

\def\lta{\mathrel{\spose{\lower 3pt\hbox{$\mathchar"218$}}
     \raise 2.0pt\hbox{$\mathchar"13C$}}}
\def\gta{\mathrel{\spose{\lower 3pt\hbox{$\mathchar"218$}}
     \raise 2.0pt\hbox{$\mathchar"13E$}}}
\newcommand{\be}{\begin{equation}}
\newcommand{\en}{\end{equation}}
\newcommand{\bea}{\begin{eqnarray}}
\newcommand{\ena}{\end{eqnarray}}
\newcommand{\ex}{\mbox{e}}
\newcommand{\dd}{\mbox{d}}

\def\setR{\mathbb{R}}
\def\setC{\mathbb{C}}

\newcommand{\ie}{\textsl{i.e.~}}

\newcommand{\eg}{\textsl{e.g.~}}
\newcommand{\etal}{\textsl{et al.~}}

\newcommand{\Hu}{{\cal H}} 
 
\newcommand{\GN}{G_{_\mathrm{N}}}

\newcommand{\lP}{\ell_{_\mathrm{Pl}}}
\newcommand{\GReCO}{${\cal G}\setR\varepsilon\setC{\cal O}$}

\title{Tensor Perturbations in Quantum Cosmological Backgrounds}

\author{Patrick Peter\\ Institut d'Astrophysique de Paris -- \GReCO,
\\ UMR CNRS, Universit\'e Pierre et Marie Curie, 98bis boulevard Arago,
\\ 75014 Paris, France. \\ E-mail: \email{peter@iap.fr} }

\author{Emanuel Pinho\\ Lafex - Centro Brasileiro de Pesquisas
F\'{\i}sicas, \\ Rue Dr.  Xavier Sigaud 150, Urca 22290-180 \\ Rio de
Janeiro, RJ, Brazil.\\ E-mail: \email{emanuel@cbpf.br}}

\author{Nelson Pinto-Neto\\ Lafex - Centro Brasileiro de Pesquisas
F\'{\i}sicas, \\ Rue Dr.  Xavier Sigaud 150, Urca 22290-180 \\ Rio de
Janeiro, RJ, Brazil. \\ E-mail: \email{nelsonpn@cbpf.br}}

\abstract{In the description of the dynamics of tensor perturbations
on a homogeneous and isotropic background cosmological model, it is
well known that a simple Hamiltonian can be obtained if one assumes
that the background metric satisfies Einstein classical field
equations. This makes it possible to analyze the quantum evolution of
the perturbations since their dynamics depends only on this classical
background. In this paper, we show that this simple Hamiltonian can
also be obtained from the Einstein-Hilbert lagrangian without making
use of any assumption about the dynamics of the background metric. In
particular, it can be used in situations where the background metric
is also quantized, hence providing a substantial simplification over
the direct approach originally developed by Halliwell and Hawking.}


\begin{document}

\section{Introduction}

In the theory of the evolution of cosmological perturbations, simple
equations have been obtained using the assumption that the background
model satisfies classical General Relativity. Lagrangians (and
Hamiltonians) describing the dynamics of scalar, vector, and tensor
perturbations coming from the Einstein-Hilbert lagrangian have been
greatly simplified in different cosmological scenarios under the
assumption that the background metric satisfies Einstein classical
field equations, after taking out space and time total
derivatives~\cite{MFB}. In such a framework, the quantization of these
perturbations becomes easy, with a quite simple interpretation: they
can be seen as quantum fields which behave essentially as scalar
fields with a time dependent effective mass. The time varying scale
factor which is responsible for this ``mass'' acts as a pump
field~\cite{grish}, creating or destroying modes of the
perturbations. Under the assumption of an initial vacuum state, the
spectrum of perturbations can be obtained and compared with
observations. This was done in particular in the cosmological
inflationary scenario~\cite{inflation}, with a resulting spectrum for
scalar perturbations in good agreement with the data~\cite{WMAP}.

As a step forward, and as the overwhelming majority of classical
backgrounds possess an initial singularity at which the classical
theory is expected to break down, it is important to study the quantum
evolution of the perturbations when the background is also quantized.
In recent years, many such quantum background cosmological models have
been proposed which share this attractive property of exhibiting
neither singularities nor horizons~\cite{pinto,pinto2,fabris}.
However, the usual treatment for cosmological perturbations in quantum
cosmological backgrounds, \ie taking the Einstein-Hilbert lagrangian
and expanding it up to second order without using the classical
background solution, in general yields extremely complicated
Hamiltonians, and, consequently, quantum equations that are difficult
to handle~\cite{halli}.

The aim of this paper is to show that these complicated equations can
be transformed into much simpler equations, at least in the case of
tensor perturbations, similar to the equations present in the
classical General Relativity case, without making any assumption about
the background dynamics. As the matter fields model we take a general
perfect fluid, for which several simple quantum cosmological solutions
at zeroth order~\cite{pinto,fabris} are known, and we briefly discuss,
for the sake of completeness, the case of matter being in the form of
a scalar field.

In order to achieve our goal, in the Hamiltonian point of view, we
show that, without ever using the background equations, there are
classical and quantum canonical transformations (see, \eg
Ref.~\cite{ander}), which we exhibit explicitly, connecting the
original Hamiltonian to the simpler one that would have been obtained
had we used the background classical field equations and,
consequently, a much simpler quantum mechanical functional
equation. In the lagrangian point of view, we eliminate a total time
derivative from the Einstein-Hilbert action, obtaining a lagrangian
with a second order derivative of the scale factor. We then use
Ostrogradski method~\cite{ostro} to handle these second derivative
terms in the lagrangian. Using the theory of constrained systems to
deal with the second class constraints that subsequently appear, we
recover the same simple Hamiltonian obtained through canonical
transformations.

The Hamiltonian can be further simplified in the case for which the
background is classical through a time dependent canonical
transformation where the scale factor is viewed as a given function of
time. When the background is also quantized, this further step can
only be implemented provided one uses an ontological interpretation of
quantum mechanics, whereby quantum Bohmian trajectories at zeroth
order can be obtained and given a meaning. Then, and only in this
case, can the ``pump field'' be treated in the gravitational wave
quantum equations as a given function of time.

The paper is organized as follows. In the following section
\ref{sec:flrw}, we specify the action and Hamiltonian by restricting
attention to the particular case of a
Friedmann-Lema\^{\i}tre-Robertson-Walker (FLRW) background over which
we concentrate on tensor perturbations only. This is done in the case
of a perfect fluid, and provides a summary of the general formalism
first obtained in Ref.~\cite{halli}. Then, after taking out a total
time derivative, we assume that the background metric satisfies the
zeroth order Einstein equations in order to simplify the action, thus
getting the simple Hamiltonian of Ref.~\cite{MFB}. The core of this
paper is then presented in Sec.~\ref{sec:qbck}, in which we show how
to obtain the simplified Hamiltonian without ever using that the
background metric satisfies the zeroth order Einstein equations, in
fact, without assuming any background dynamics, both in the
Hamiltonian and Lagrangian point of views. In Sec.~\ref{sec:further},
we show how further simplifications can be achieved if one assumes the
Bohm-de Broglie interpretation.  This is also where we briefly discuss
the scalar field case. Finally, Sec.~\ref{sec:conc} ends this paper
with some general conclusions.

\section{Review of Tensor Perturbations in a Classical
Friedmann Lema\^{\i}tre Robertson Walker Background}
\label{sec:flrw}

In this section we review the procedures to obtain the Hamiltonian
governing the classical and quantum dynamics of tensor perturbations
when one assumes that the cosmological background satisfies classical
Einstein's equations.

\subsection{General action and Hamiltonian}

Let us consider the case in which the background spacetime is of the
Friedmann-Lema\^{\i}tre-Robertson-Walker (FLRW) type, over which we
wish to investigate tensor perturbations. As it is well known, this
spacetime may be foliated by space-like hypersurfaces which are
maximally symmetric, and we shall also impose that these hypersurfaces
be compact. Therefore, the line element may be written as
\begin{equation}
\dd s^2=\dd t^2- a^2(t)\gamma_{ij} \dd x^{i}\dd x^{j},
\label{tf}
\end{equation}
where $t$ is the physical time as measured by an observer co-moving
with the maximally symmetric hypersurfaces, $a(t)$ is the so called
scale factor and $\gamma_{ij}$ is the metric of the spacelike
hypersurface.

This line element may be alternatively written as
\begin{equation}
\dd s^2= a^2(\eta) \left( \dd \eta^2- \gamma_{ij}\dd x^{i}\dd
x^{j} \right),
\label{tc}
\end{equation}
where $\eta$ is called the conformal time. We shall in what follows
use the more general expression
\begin{equation}
\dd s^2= N^2(\tau)\dd\tau^2- a^2(\tau)\gamma_{ij} \dd x^{i} \dd x^{j}.
\label{adm}
\end{equation}
Both expressions (\ref{tf}) and (\ref{tc}) then follow from
(\ref{adm}) by adequate choices of the function $N$ and concomitantly
of the timelike coordinate $\tau$.  Specifically, (\ref{tc}) is
equivalent to $N=a$. In the ADM formalism, $N$ is the lapse
function. For FLRW spacetimes the shift vector may be made identically
zero.

We shall now perturb the above geometry by adding to the metric the
infinitesimal quantities $w_{ij}$ in the following manner:
\begin{equation}
\dd s^{2}=N^2\left(\tau\right)\dd\tau^2
-a^2\left(\tau\right) \left(\gamma_{ij}+w_{ij}\right)\dd x^{i}\dd x^{j},
\label{adm2}
\end{equation}
which transform as a tensor under diffeomorphisms of the 3-surface
onto itself; $w_{ij}$ satisfy the identities
\begin{equation} w^{ij}\,_{|i}=0, \qquad \hbox{and} \qquad
w^{i}\,_{i}=0. \label{twz}
\end{equation}
In the above expression, the bar stands for the covariant derivative
using as connections the Christoffel symbols calculated from
$\gamma_{ij}$, and indices of $w_{ij}$ are raised and lowered by this
metric. It is the purpose of the present article to study the
dynamics of the tensor modes $w_{ij}$, which are the ones that give
rise to gravitational waves on the FLRW background. It is also
possible to show that these tensor modes are gauge invariant and
therefore can never be coordinate artifacts, having a clear physical
interpretation as real perturbations.

Let us now suppose that the FLRW background is filled with a fluid
that obeys the equation of state
\begin{equation}
p=(\lambda - 1)\rho,
\end{equation}
$p$ and $\rho$ being respectively the pressure and energy density of
the fluid, and $\lambda$ being a constant. Note that all the following
results can easily be generalized to any matter field minimally
coupled to gravity. The example of a minimally coupled scalar field is
actually treated in section~\ref{sec:scalarfield}. Here, we chose a
perfect fluid because there are some known (and simple) minisuperspace
quantum solutions in the literature~\cite{pinto,fabris} in that case,
which can be immediately applied to the following results.  If we
restrict ourselves to the case where only the tensor modes are
present, we can show, using the background FLRW metric given in
(\ref{adm}), that the gravitational part of the action, namely
\begin{equation}
\mathcal{S}_{_\mathrm{GR}} = -\frac{1}{6\lP^2} \int
\sqrt{-g} R \dd^4 x, \label{ActionGR}
\end{equation} 
[$\lP=(8\pi\GN/3)^{1/2}$ being the Planck length] may be written as
\begin{eqnarray}
\mathcal{S}_{_\mathrm{GR}}=-\frac{1}{6\lP^2} \int \dd^{4} x \,
\gamma^{1/2} a^{3}& &\bigl[\frac{6\Hu^2}{N}-\frac{6Nk}{a^2} -
\frac{\dot{w}^{ij}\dot{w}_{ij}}{4N} + \frac{Nw^{ij|k}w_{ij|k}}{4a^2}
\cr & & -
\left(\frac{3\Hu^2}{2N} - \frac{kN}{a^2} \right) w_{ij}
w^{ij} -
\frac{2\Hu w^{ij}\dot{w}_{ij}}{N}\bigr] ,
\label{graction}
\end{eqnarray}
where ${\cal H}\equiv\dot{a}/a$ and the dot means derivative with
respect to coordinate time $\tau$. 

For the matter part of the action, the fluid has as lagrangian density
\begin{equation}
\mathcal{L}_\mathrm{matter} = p \sqrt{-g} \label{l}
\end{equation}
As the tensor modes are the only perturbations present, and as they
obey Eqs.~(\ref{twz}), no combination of them can appear in
(\ref{l}). Conversely, by the assumed linearity of the dynamics of the
perturbations, no perturbation of $p$, which is a scalar, can couple
to the tensor mode under study. Hence, the pressure and energy density
of the fluid will be kept to their background values. Therefore, the
matter part of the action changes according to
\begin{equation}\delta^{(2)}\mathcal{S}_\mathrm{matter} = \int \dd^4 x
\, \gamma^{1/2} Na^{3} \left( 1 - \frac{1}{4} w^{ij} w_{ij} \right) p.
\end{equation}
The total action reads
\begin{eqnarray}
\mathcal{S}&=&-\int \frac{\dd^{4} x}{6\lP^2}  \,
\gamma^{1/2}a^{3}\left[ \frac{6\Hu^2}{N} - \frac{6Nk}{a^2} -
\frac{\dot{w}^{ij}\dot{w}_{ij}}{4N} +
\frac{Nw^{ij|k}w_{ij|k}}{4a^2}-\left(\frac{3\Hu^2}{2N} -
\frac{kN}{a^2}\right) w_{ij} w^{ij} \right. \cr & & \left.- \frac{2\Hu
w^{ij} \dot{w}_{ij}}{N} \right] + \int \dd^{4} x \,
\gamma^{1/2}a^{3} N p \left( 1 + \frac{w_{ij} w^{ij}}{4} \right).
\label{at1}
\end{eqnarray}
The pure gravitational part of this action is the one that appears in
Ref.~\cite{halli}, where the tensorial modes are decomposed in terms
of the normal tensor modes, which are eigenfunctions of the Laplacian
operator on the 3-sphere (in that article, the analysis was restricted
to the case $k=+1$ and the fluid is a non-minimally coupled scalar
field). Of course, it generates Einstein's equations at zeroth and
first order in the tensor perturbation, as can be checked.

We can now calculate the Hamiltonian coming from action
(\ref{at1}). Let us first recall the expressions for the canonical
momenta, namely
\begin{equation}
P_a = \frac{1}{6\lP^2} \int \dd^3 x \, \gamma^{1/2} \frac{a^2}{N}
\left( -12\Hu+3\Hu w_{ij} w^{ij} + 2 w_{ij}\dot{w}^{ij}\right),
\end{equation}
and
\begin{equation}
\tilde{\Pi}^{ij} = \frac{\gamma^{1/2}a^3}{6\lP^2 N}\left( 2\Hu w^{ij}
+ \frac{1}{2}\dot{w}^{ij}\right).
\label{pp}
\end{equation}
The Hamiltonian can then be calculated noting that $a$ and $P_a$
depend only on the parameter $\tau$, and therefore can be taken
outside the spatial integrals. In what follows, we also define the
total volume of three-space through $V\equiv \int \dd^3
x\,\gamma^{1/2}$, which we assume is finite, \ie, as previously
discussed, we consider either closed spatial sections, or compact flat
or hyperbolic sections.

For the fluid part, we use the formalism of Schutz \cite{schutz},
in which the pressure $p$ of the fluid is written as
\begin{equation}
p= p_{0r}\left( \frac{\dot{\varphi}+\theta\dot{s}}{N\lambda}
\right)^{\frac{\lambda}{\lambda-1}}
\exp{\left[-\frac{s}{s_{0r}\left(\lambda-1\right)} \right]}.
\end{equation}
The variables $\varphi$, $\theta$ and $s$ are velocity potentials for
the fluid congruence with suitable thermodynamical interpretations,
and $p_{0r}$ and $s_{0r}$ are arbitrary constants related to the
initial conditions of the fluid. The canonical momenta $p_{\varphi}$,
$p_{s}$ and $p_{\theta}$ can be obtained in the usual way. One can
perform the canonical transformation (for details see
Refs.~\cite{pinto2,russo})
\begin{equation}
T=-p_s\exp \left( -\frac{s}{s_{0r}}\right) p_\varphi^{-\lambda}
\rho_{0r}^{\lambda-1}s_{0r},
\label{can1}
\end{equation}
and
\begin{equation}
\varphi_N=\varphi + \lambda s_0 \frac{p_s}{p_\varphi},
\label{can3}
\end{equation}
leading to the momenta
\begin{equation}p_{_T}=\frac{p_\varphi^\lambda}{\rho_{0r}^{\lambda-1}}
\exp\left(-\frac{s}{s_{0r}}\right),
\label{can2}
\end{equation}
and
\begin{equation}
p_{\varphi_N}=p_{\varphi}.
\label{can4}
\end{equation}
As it turns out, these variables are more suitable than the original
ones as the fluid Hamiltonian expressed in terms of those gets a
much simpler form.

We can also rescale the variables $a$, $w$, $N$, $T$, and their
momenta, according to
\begin{eqnarray}\bar{a}&=&a\frac{\sqrt{V}}{\lP}, \label{red1} \\
\bar{P}_a&=& P_a\frac{\lP}{\sqrt{V}},\\
\bar{w}_{ij}&=&\frac{w_{ij}}{\sqrt{V}},\\
\bar{\tilde{\Pi}}\null^{ij}&=&\tilde{\Pi}\null^{ij}\sqrt{V},\\
\bar{T}&=&TV,\\ \bar{p}_{_T}&=&\frac{p_{_T}}{V},\\
\bar{N}&=&N\frac{\sqrt{V}}{\lP},\label{red2}
\end{eqnarray}
and a dimensionless Hamiltonian
\begin{equation}\bar{H}=H\frac{\lP}{\sqrt{V}}.\end{equation}
After all these procedures, the total Hamiltonian is finally put in
the form (omitting from now on the bars):
\begin{eqnarray}H&\equiv&NH_0\nonumber \\
&=&N\left\{-\frac{P_a^{2}}{4a}-ka+ \frac{P_{_T}}{a^{3(\lambda-1)}}
\left[1+\displaystyle\frac{\left(\lambda -1\right)}{4} \int \dd^{3} x
\gamma^{1/2}\, w_{ij} w^{ij}\right] \right. \cr
&+& \frac{5P_a^2}{48a} \int \dd^3 x
\gamma^{1/2}\, w_{ij}w^{ij} \nonumber\\&+& \!\!\left. \int \dd^3
x \left[ \frac{6P_{(w)ij} {\tilde{\Pi}}\null^{ij}}{a^{3} \gamma^{1/2}}
+ 2\frac{P_a w_{ij}\tilde{\Pi}^{ij}}{a^2} + \gamma^{1/2}a \left(
\frac{w^{ij|k} w_{ij|k} }{24} + \frac{k}{6} w_{ij} w^{ij}
\right)\right] \right\}\!\!,
\label{h1}
\end{eqnarray}
which is nothing but the Hamiltonian of Ref.~\cite{halli} expressed
for a perfect fluid. This Hamiltonian, which is zero due to the
constraint $H_0\approx0$, yields the correct Einstein equations both
at zeroth and first order in the perturbations, as can be checked
explicitly. Note that in order to obtain its expression, no assumption
has been made about the background dynamics.

The first order equation for the tensor perturbation reads
\begin{equation}
\label{weq}
\ddot{w}_{ij} - \frac{\dot{N}}{N} \dot{w}_{ij} + 3\frac{\dot{a}}{a}
\dot{w}_{ij} - \frac{N^2}{a^2} w^{|l}_{ij|l} + 2 k\frac{N^2}{a^2}
w_{ij}=0.
\end{equation}

We will show in the next section that if one assumes that the
background satisfies separately Einstein classical field equations,
one may achieve a considerable simplification of the Hamiltonian which
leads to the quantum equations for the tensor perturbations, which in
this case are the sole degrees of freedom to be quantized.

\subsection{Classical and quantum gravitational waves on a classical
  FLRW background}
\label{subsec:qgw}

We will now assume that the background spacetime and its matter
content are well described by classical GR in order to obtain a
simpler Hamiltonian than Eq.~(\ref{h1}), and a simpler quantum
dynamics for the gravitational waves.

Varying the action (\ref{at1}) with respect to $a(\tau)$, and keeping
only the terms up to zeroth order, we obtain the following equation of
motion

\begin{equation}
\frac{\dot{N}}{N} \Hu - \frac{\ddot{a}}{a} - \frac{\Hu^2}{2} -
\frac{kN^2}{2a^2} - \frac{3\lP^2}{2} p N^2 = 0,
\label{em}
\end{equation}
which is nothing but one of the Friedmann's equations. On the other
hand, integrating by parts the term $w_{ij}\dot{w}^{ij}$ in the action
(\ref{at1}) yields
\begin{eqnarray}
\mathcal{S}&=&\frac{1}{6\lP^2} \int \dd^4 x \gamma^{1/2}\, a^{3}
\left[ \frac{1}{N} \left( \frac{\dot{N}}{N} \Hu - \frac{\Hu^2}{2} -
\frac{\ddot{a}}{a} - \frac{3\lP^2}{2} p N^2 - \frac{N^2 k}{a^2}\right)
w_{ij} w^{ij}\right.\nonumber \\ & & \left. + \frac{1}{4N}
\dot{w}_{ij} \dot{w}^{ij} - \frac{N}{4a^2} w_{ij|k} w^{ij|k} -
\frac{6\Hu^2}{N} + \frac{6Nk}{a^2} + 6\lP^2 N p\right],
\label{ai}
\end{eqnarray}
where a total time derivative term has been omitted. Using
Eq.~(\ref{em}), one gets
\begin{eqnarray}
\mathcal{S}&=&\frac{1}{6\lP^2} \int \dd^4 x \gamma^{1/2}\,a^3 \left( -
\frac{6\Hu^2}{N} + 6\frac{Nk}{a^2} + 6\lP^2 N p \right. \cr
& &\ \ \ + \left.\frac{1}{4N}
\dot{w}_{ij} \dot{w}^{ij} - \frac{N}{4a^2} w_{ij|k} w^{ij|k} -
\frac{Nk}{2a^2} w_{ij} w^{ij} \right).
\end{eqnarray}
The second order terms correspond to the action for gravitational
waves on a FLRW background as presented in Ref.~\cite{MFB} (shown
there in the gauge $N=a$),
\begin{equation}
S_{_\mathrm{GW}} = \frac{1}{6\lP^2} \int \dd^4 x \gamma^{1/2}\,a^{3}
\left( \frac{1}{4N} \dot{w}_{ij} \dot{w}^{ij} - \frac{N}{4a^2}
w_{ij|k} w^{ij|k} - \frac{Nk}{2a^2} w_{ij} w^{ij}\right).
\end{equation}

Calculating the momenta canonically conjugate to the variables $a$ and
$w_{ij}$ then yields
\begin{equation}
P_a=-\frac{2\Hu a^2 V}{\lP^2 N},
\end{equation}
and
\begin{equation}
\Pi^{ij} = \frac{\gamma^{1/2}a^{3}\dot{w}^{ij}}{12\lP^2 N}.
\end{equation}
Inverting these relations, and making the canonical transformations
(\ref{can1})~--~(\ref{can4}) in the fluid sector, as well as the
rescaling (\ref{red1})~--~(\ref{red2}), we obtain for the Hamiltonian
of the classical FLRW background and gravitational waves,
\begin{eqnarray}
H&=&N\left[ -\frac{P_a^2}{4a} - ka+ \frac{P_{_T}}{a^{3(\lambda - 1)}}
\right.\cr & &\ \ \left. + \int \dd^3 x \left(
6\frac{\Pi^{ij}\Pi_{ij}}{\gamma^{1/2}a^3} + \frac{1}{24} \gamma^{1/2}
a w_{ij|k} w^{ij|k}+ \frac{1}{12} \gamma^{1/2} k
w_{ij}w^{ij}a\right)\right].
\label{h2}
\end{eqnarray}
This Hamiltonian also yields the correct Einstein equations at zeroth
and first order in the perturbations, in particular Eq.~(\ref{weq}).
Besides, and because only tensor perturbations will be quantized in
what follows, we shall focus on the gravitational wave part of
Hamiltonian (\ref{red2}), namely
\begin{equation} H_\mathrm{gw} = \frac{N}{a}
H^\mathrm{c}_\mathrm{gw} = \int \dd^3 x \left( 6
\frac{\Pi^{ij}\Pi_{ij}}{\gamma^{1/2} a^{2}} + \frac{1}{24}
\gamma^{1/2} a^2 w_{ij|k}w^{ij|k} + \frac{1}{12} \gamma^{1/2} k w_{ij}
w^{ij} a^2\right).
\label{hgw2}
\end{equation}
The transformation achieved here already provides an important
simplification over the original Hamiltonian~(\ref{h1}), but, as we
shall now see, one can go even further.

As, for the moment, $a$ is not quantized and satisfies the background
classical GR equations, one can view it simply as a function of the
time parameter present in the Hamiltonian $H_\mathrm{gw}$. One can
then put $H_\mathrm{gw}$ in an even simpler and suggestive form by
performing the following time dependent canonical transformation
\begin{equation}
w_{ij}=\frac{\sqrt{12}}{a} \mu_{ij}, \qquad \hbox{and} \qquad
\Pi^{ij}=\frac{1}{\sqrt{12}} \left( a\Pi_{(\mu)}^{ij} - \gamma^{1/2}
\dot{a} \mu^{ij}\right),
\label{ct2}
\end{equation}
whose generating functional is
\begin{equation}
F_2\left[ w_{ij},\Pi_{(\mu)}^{ij},\tau\right] = 
\frac{1}{\sqrt{12}}\int \dd^3 x \left( a w_{ij} \Pi_{(\mu)}^{ij} -
\gamma^{1/2} \frac{a \dot{a} w_{ij} w^{ij}}{2\sqrt{12}}\right),
\label{f2}
\end{equation}
from which one obtain the transformation itself through the equations
\begin{equation} \mu_{ij} = \frac{\delta F_2}{\delta\Pi_{(\mu)}^{ij}},
\qquad \hbox{and} \qquad \Pi^{ij} = \frac{\delta F_2}{\delta w_{ij}}.
\end{equation}
The new Hamiltonian will be given by
\begin{equation}
\tilde{H}_\mathrm{gw} = H_\mathrm{gw} + \frac{\partial F_2}{\partial
\tau}, 
\end{equation}
and it reads
\begin{equation}
\tilde{H}_\mathrm{gw} = \frac{N}{a} \tilde{H}^\mathrm{c}_\mathrm{gw}
= \int \dd^3 x \left[ \frac{\Pi_{(\mu)}^{ij}
\Pi_{(\mu)ij}}{2\gamma^{1/2}} + \frac{1}{2}\gamma^{1/2}\mu_{ij|k}
\mu^{ij|k} + \gamma^{1/2}\left( k-\frac{\ddot{a}}{2a} \right) \mu_{ij}
\mu^{ij} \right].
\label{hgw3}
\end{equation}
In the conformal gauge $N=a$ (for which $\tau=\eta$), the Hamiltonian
(\ref{hgw3}) yields the following equation for $\mu$:
\begin{equation}
\label{weqmu}
{\ddot{\mu}}_{ij}-\mu^{|l}_{ij|l}+\left(2k- \frac{\ddot{a}}{a}
\right) \mu_{ij}=0,
\end{equation}
which can be obtained from Eq.~(\ref{weq}) by making the substitutions
$N=a$ and $w_{ij}=\mu_{ij}/(\sqrt{12}a)$ in this relation.

The classical background geometry furnishes the time parameter (in the
gauge chosen, the conformal time) on which the wave functional
$\psi(\mu_{ij},\eta)$ evolves. The functional Schr\"odinger equation
it satisfies, namely
\begin{equation}
i\frac{\partial|\psi\rangle}{\partial \eta}=
\hat{\tilde{H}}\null^\mathrm{c}_\mathrm{gw}|\psi\rangle,
\label{schro0}
\end{equation}
reads explicitly, in the coordinate representation,
\begin{eqnarray}
\label{schro1}
i\frac{\partial\psi(\mu_{ij},\eta)}{\partial \eta}= &\int &\dd^3 x
\left\{ - \frac{1}{2\gamma^{1/2}} \frac{\delta^{2}}{\delta \mu_{ij}
\delta \mu^{ij}} + \gamma^{1/2}\left[ \frac{1}{2} \mu_{ij|k}
\mu^{ij|k} + \left( k- \frac{\ddot{a}}{2a} \right) \mu_{ij} \mu^{ij}
\right] \right\} \cr \cr
&\times& \psi(\mu_{ij},\eta)=0.
\end{eqnarray}

One can also show that quantization in the gauge $N=a$ of
gravitational waves using the Hamiltonian $H^\mathrm{c}_\mathrm{gw}$
defined in Eq.~(\ref{hgw2}), through the Schr\"odinger equation
\begin{equation}
i\frac{\partial|\varphi\rangle}{\partial \eta}=
\hat{H}^\mathrm{c}_\mathrm{gw}|\varphi\rangle,
\label{schro2}
\end{equation}
yields an equivalent quantum theory as the one described by
Eq.~(\ref{schro1}). In fact, there is a time dependent quantum
canonical transformation mapping the two theories, generated by the
unitary operator (for a good review on this subject, see
Ref.~\cite{ander}),
\begin{equation}
\label{qct0}
U = \exp\left\{ i \left[ \int \dd^3 x \gamma^{1/2} \frac{\dot{a}
w_{ij} w^{ij}}{2a} \right] \right\} \exp\left\{ i \left[ \int \dd^3 x
\left( \frac{w_{ij}\Pi ^{ij} + \Pi ^{ij} w_{ij}}{2} \right) \ln\left(
\frac{\sqrt{12}}{a} \right) \right]\right\}.
\end{equation}

Under the map between self-adjoint operators
\begin{equation}
\hat{\tilde{A}}= U\hat{A}U^{-1},
\end{equation}
one can obtain the operator version of Eq.~(\ref{ct2}), making use in
the derivation, of the Baker -- Campbell -- Hausdorff formula
\begin{equation}
\ex^A B \ex^{-A} = B + [A,B] +\frac{1}{2!} [A,[A,B]]+ \cdots
\label{BCH}
\end{equation}
The operator $\hat{\tilde{H}}\null^\mathrm{c}_\mathrm{gw}$ is obtained
from $\hat{H}^\mathrm{c}_\mathrm{gw}$ through the usual formula when
$U$ is time dependent, namely
\begin{equation}
\hat{\tilde{H}}\null^\mathrm{c}_\mathrm{gw} = U
\hat{H}^\mathrm{c}_\mathrm{gw} U^{-1} +
i \frac{\partial U}{\partial \eta} U^{-1}.
\end{equation}
One can show that if $|\varphi\rangle$ is a solution of
Eq.~(\ref{schro2}), then $|\psi\rangle=U|\varphi\rangle$ is a solution
of Eq.~(\ref{schro0}) and all probability amplitudes of the two
theories are equal as long as $U$ is unitary (the operators in the
exponentials are self-adjoint):
$\langle\varphi_1|\hat{A}|\varphi_2\rangle=\langle \psi_1
|U\hat{A}U^{-1}|\psi_2\rangle=\langle \psi_1
|\hat{\tilde{A}}|\psi_2\rangle$.

The Schr\"odinger equation (\ref{schro1}) is evidently an enormous
simplification, made in two steps, over the one that would have been
obtained if one had sticked to the Hamiltonian (\ref{h1}), with a
quite simple interpretation: it describes a scalar field with time
dependent mass given by $-\ddot{a}/a$. This time dependence is
responsible for the creation and/or annihilation of tensor modes due
to the pump field $a$ governing the dynamics of the cosmological
background.

The consequences of Eq.~(\ref{schro1}) for tensor perturbations in
various classical background models have been extensively discussed in
the literature (see \eg Refs.~\cite{MFB,grish,outros}). We will now
investigate the situation where the background is also quantized.

\section{Quantum Gravitational Waves on a Quantum FLRW Background}
\label{sec:qbck}

The two Hamiltonians (\ref{h1}) and (\ref{h2}) are completely
equivalent at the classical level, and (\ref{h2}) can be obtained from
(\ref{h1}) if one assumes that the background satisfies classical
GR. However, if the background is also quantized, that is, when $a$
becomes an operator, one can obviously no longer use this method to
obtain (\ref{h2}) from (\ref{h1}). Indeed, in Eq.~(\ref{h2}) $a$ is
assumed to be that prescribed function of time (not a canonical
variable) which satisfies the classical equations of motion. This is
incompatible with $a$ viewed as a quantum operator, coming from the
quantization of the canonical variable $a$. Hence, one may question
the use of equation (\ref{h2}) as a valid Hamiltonian when the
background is also quantized.

It is our goal in this section to show that equation (\ref{h2}), with
$a$ viewed as a {\it canonical variable} and {\it not} as {\it a
prescribed function of time}, can indeed be obtained from (\ref{h1})
{\it without ever making any assumption concerning the dynamics of the
background}, and that the simpler Hamiltonian (\ref{h2}) can be used
in the canonical quantization of the whole system. This equivalence is
also proved at the quantum level. Indeed, simplifying the Hamiltonian
constraint is very important at the quantum level since the Dirac
canonical quantization procedure for constrained systems is obtained
by imposing that the physical states $\Psi(a,w_{ij},t)$ are
annihilated by the operator version of $H_0$, \ie, ${\hat{H}}_0 \Psi
=0$.  Using the operator version of Eq.~(\ref{h2}) happens to be much
simpler than that arising from (\ref{h1}), which is rather complicated
and also suffers from many factor ordering ambiguities.

\subsection{The Hamiltonian point of view: classical and quantum
canonical transformations} \label{sec:Hclq}

The difference between actions (\ref{at1}) and (\ref{ai}) is a total
time derivative given by,
\begin{equation}
\Delta \mathcal{S}= -\int \dd^4 x \frac{\dd}{\dd t}\left( \gamma^{1/2}
\frac{\dot{a} a^2 w_{ij} w^{ij}}{6\lP^2 N}\right)= \int \dd^4 x
\frac{\dd}{\dd t}\left( \gamma^{1/2} \frac{P_a a w_{ij}
w^{ij}}{12\lP^2}\right),
\end{equation}
which suggests that, after making the redefinitions
(\ref{red1})~--~(\ref{red2}), a canonical transformation generated by
\begin{equation}
G = a \tilde{P}_a - \int \dd^3 x\, \tilde{w}_{ij} \Pi^{ij}
+ \int \dd^3 x \gamma^{1/2} \, \frac{\tilde{P}_a a \tilde{w}_{ij}
\tilde{w}^{ij}}{12},
\end{equation}
will transform (\ref{h1}) into (\ref{h2}) (the first two terms
yielding the identity part of the transformation). Indeed, through the
relations [here, the tilde variables refer to Hamiltonian (\ref{h1})]
\begin{eqnarray}
\tilde{a} & = & \frac{\partial G}{\partial\tilde{P}_a},\\ & &
\nonumber \\ P_a & = & \frac{\partial G}{\partial a}, \\ w_{ij} & = &
-\frac{\delta G}{\delta \Pi^{ij}},\\ \Pi^{ij} & = & -\frac{\delta
G}{\delta \tilde{w}_{ij}},
\end{eqnarray}
we obtain, up to second order in the perturbation
\begin{eqnarray}
\tilde{a} & = & a\left( 1 + \frac{Q}{12}\right),\label{tc20a} \\
\tilde{P}_a & = & P_a \left( 1 - \frac{Q}{12}\right), \\
\tilde{w}_{ij}&=& w_{ij},\\ \tilde{\Pi}^{ij}&=&\Pi^{ij}-
\frac{1}{6}\gamma^{1/2}a P_a w^{ij},
\label{tc20}
\end{eqnarray}
where we have defined
\begin{equation}
Q\equiv\int\dd^3 x \gamma^{1/2}\, w_{ij}w^{ij}.
\label{defQ}
\end{equation}
Note that the equation (\ref{tc20}) is the same as Eq.~(\ref{pp}).
One can easily check that Eqs.~(\ref{tc20a}) to (\ref{tc20}) are
canonical transformations (up to second order) which transforms
(\ref{h1}) into (\ref{h2}) (also up to second order),
\begin{eqnarray}
H=N&&\left[ -\frac{P_a^2}{4a} - ka+ \frac{P_{_T}}{a^{3(\lambda - 1)}}
\right. \cr &&\ \ \ \left. + \int \dd^3 x \left(
6\frac{\Pi^{ij}\Pi_{ij}}{\gamma^{1/2}a^3} + \frac{1}{24} \gamma^{1/2}
a w_{ij|k} w^{ij|k}+ \frac{1}{12} \gamma^{1/2} k
w_{ij}w^{ij}a\right)\right].
\label{h207}
\end{eqnarray}
Eq.~(\ref{h207}) has the same simple form as Eq.~(\ref{h2}), but here
$a$ and $P_a$ are canonical variables like $w_{ij}$ and $\Pi^{ij}$
ready to be quantized.

Quantum mechanically, the above canonical transformations are
generated by the unitary operator
\begin{equation}
\label{u2}
U=\exp(iG_{\rm q})
\equiv\exp\left(\frac{i}{12}\hat{\beta}_a \hat{Q}\right),
\end{equation}
where $\hat{\beta}_a \equiv (\hat{P}_a \hat{a} + \hat{a}\hat{P}_a)/2$
and $\hat{Q} \equiv \int \dd^3 x\, \gamma^{1/2} \hat{w}_{ij}
\hat{w}^{ij}$ are the self-adjoint operators associated with the
corresponding classical variables.

In a way completely analogous to that of the previous section, we
perform the map $\hat{A}\mapsto\hat{\tilde{A}}= U\hat{A}U^{-1}$ for
all operators $\hat{A}$ in such a way as to obtain the operator
version of Eq.~(\ref{tc20}), making, here again, use of relation
(\ref{BCH}). Note that the map generated by (\ref{u2}) yields
$\hat{\tilde{a}}=\hat{a}\exp(Q/12)$ and $\hat{\tilde{P}}_a=\hat{P}_a
\exp(-Q/12)$, which reduces to the operator version of
Eq.~(\ref{tc20}) up to second order, with the factor ordering for the
transformation of $\hat{{\tilde{\Pi}}}\null^{ij}$ given by
$\hat{{\tilde{\Pi}}}\null^{ij}=\hat{\Pi}\null^{ij}- \gamma^{1/2}
\hat{\beta}_a \hat{w}^{ij}/6$.

The transformation of the operator version of (\ref{h1}) into the
operator version of (\ref{h2}) depends on the factor ordering of the
operator version of (\ref{h1}). For instance, for the ordering
\begin{eqnarray}
\hat{H}_0 & = & -\frac{\hat{\beta}_a^2}{4\hat{a}^3} - k \hat{a} +
\frac{\hat{P}_{_T}}{\hat{a}^{3(\lambda-1)}} \left[ 1 + \frac{\left(
\lambda -1\right)}{4} \int \dd^3 x \, \gamma^{1/2} \hat{w}_{ij}
\hat{w}^{ij} \right] + \frac{5 \hat{\beta}_a^2}{48 \hat{a}^3} \int
\dd^3 x \gamma^{1/2}\hat{w}_{ij} \hat{w}^{ij} \nonumber \\ & & + \int
\dd^3 x \left[ \frac{6\hat{\Pi}_{ij} \hat{\Pi}_{}^{ij}}{\hat{a}^3
\gamma^{1/2}} + \frac{\hat{\beta}_a}{\hat{a}^3} \left( \hat{w}_{ij}
\hat{\Pi}^{ij} + \hat{\Pi}^{ij} \hat{w}_{ij}\right) \right] \cr & & +
\int \dd^3 x \gamma^{1/2}\, \hat{a} \left( \frac{\hat{w}^{ij|k}
\hat{w}_{ij|k}}{24} + \frac{k}{6} \hat{w}_{ij} \hat{w}^{ij} \right),
\cr & & \label{h10}
\end{eqnarray}
one obtains after this quantum canonical transformation
\begin{equation}
\hat{H}_0 = -\frac{\hat{\beta}_a^2}{4\hat{a}^3} - k \hat{a} +
\frac{\hat{P}_{_T}}{\hat{a}^{3(\lambda - 1)}}+ \int \dd^3 x \left( 6
\frac{\hat{\Pi}^{ij} \hat{\Pi}_{ij}}{\gamma^{1/2} a^3} +
\frac{1}{24}\gamma^{1/2} a \hat{w}_{ij|k}\hat{w}^{ij|k}+
\frac{1}{12}\gamma^{1/2} k \hat{w}_{ij} \hat{w}^{ij} a \right),
\label{h20}
\end{equation}
where the ordering of $\hat{\beta}_a^2/\hat{a}^3$
must be the same in both Hamiltonians.

Another possibility is
\begin{eqnarray}
& \hat{H}_0 & =\biggl\{ - \frac{1}{4\hat{a}}\hat{P}_a^2 - k\hat{a} +
\frac{\hat{P}_{_T}}{\hat{a}^{3(\lambda-1)}} \left[ 1 + \frac{(\lambda
-1)}{4} \int \dd^3 x \gamma^{1/2} \, \hat{w}_{ij} \hat{w}^{ij}\right]
\nonumber\\ && + \int \dd^3 x \gamma^{1/2}\, \hat{w}_{ij} \hat{w}^{ij}
\left[\frac{1}{24} \left(\hat{a}\hat{P}_a\hat{a}\hat{P}_a
\frac{1}{\hat{a}^3} + \hat{a} \hat{P}_a^2 \frac{1}{\hat{a}^2} +
\hat{P}_a \hat{a}^2 \hat{P}_a \frac{1}{\hat{a}^3} + \hat{P}_a \hat{a}
\hat{P}_a \frac{1}{\hat{a}^2}\right) \right. \cr & & \ \ \ \ \ \  \ \
\ \ \ \ \ \ \ \ \ \ \ \ \ \ \ \ \ \ \ \ \ \ \ \ \ \ \ \ \ \ \ \ \ \ \
\ \ \ \  \ \ \left. -\frac{1}{16\hat{a}}
\hat{P}_a^2\right] \nonumber\\&& + \int \dd^3 x \, \biggl[
\frac{6\hat{\Pi}_{ij} \hat{\Pi}^{ij}}{\hat{a}^3\gamma^{1/2}} +
\hat{\beta}_a (\hat{w}_{ij} \hat{\Pi}^{ij} + \hat{\Pi}^{ij}
\hat{w}_{ij} ) \frac{1}{\hat{a}^3} + \gamma^{1/2}\hat{a} \biggl(
\frac{\hat{w}^{ij|k} \hat{w}_{ij|k}}{24} + \frac{k}{6}\hat{w}_{ij}
\hat{w}^{ij} \biggr)\biggr],\cr \cr & & \null
\label{h100}
\end{eqnarray}
going to
\begin{eqnarray}
\hat{H}_0 = && \left[ -\frac{1}{4\hat{a}}\hat{P}_a^2 -k\hat{a}+
\frac{\hat{P}_{_T}}{\hat{a}^{3(\lambda - 1)}}\right. \cr && +
\left.\int \dd^3 x \left(
6\frac{\hat\Pi^{ij}\hat{\Pi}_{ij}}{\gamma^{1/2} a^3} + \frac{1}{24}
\gamma^{1/2} a \hat{w}_{ij|k}\hat{w}^{ij|k}+ \frac{1}{12} \gamma^{1/2}
k \hat{w}_{ij} \hat{w}^{ij} a \right)\right].
\label{h200}
\end{eqnarray}
Quantum evolutions governed by (\ref{h10}) and (\ref{h20}), or
(\ref{h100}) and (\ref{h200}) are equivalent. It may be noted that the
factor ordering of (\ref{h100}) and (\ref{h200}) yields non self
adjoint operators. However, as will see later on, for some fluids they
may yield Schr\"odinger like equations with appropriate self-adjoint
reduced Hamiltonian operators, and can therefore be viewed as
appropriate quantum representations.

\subsection{The Lagrangian point of view}

If we start with Eq.~(\ref{at1}), and make the same integration by
parts that led to Eq.~(\ref{ai}), we obtain an expression that
involves $\ddot{a}$ and $\dot{N}$:
\begin{eqnarray}
S & = & -\frac{V\dot{a}^2 a}{\lP^2 N} + \frac{VN}{\lP^2}\left(
ka+\lP^2 p a^3\right) \nonumber \\ & & + \frac{1}{6\lP^2} \int \dd^4 x
\gamma^{1/2}\, a^3 \left[ \frac{1}{N}\left( \frac{\dot{N}}{N}
\frac{\dot{a}}{a}- \frac{\dot{a}^{2}}{2a^{2}}-
\frac{\ddot{a}}{a}-\frac{3\lP^2}{2} p N^2-\frac{N^2 k}{a^2} \right)
w_{ij}w^{ij}\right.\cr & &\left.+ \frac{1}{4N}\dot{w}_{ij}\dot{w}^{ij}
-\frac{N}{4a^{2}}w_{ij|k}w^{ij|k}\right].
\end{eqnarray}

In the previous section, it was possible to deal with these terms by
using the classical equations of motion. As we do not want to follow
this line in this section, we consider another way and make use of the
Ostrogradski method~\cite{ostro} to write down a Hamiltonian for a
lagrangian that involves second order time derivatives. In this method
we define
\begin{equation}
b=\dot{a},
\end{equation}
and treat the variable $b$ as independent of $a$. The momenta
canonically conjugate to $a$ and $b$ are defined as
\begin{equation}
P_b=\frac{\partial L}{\partial\ddot{a}}, \quad \hbox{and}
\quad P_a = \frac{\partial L}{\partial\dot{a}}-\dot{P_b},
\end{equation}
while the other momenta maintain their usual definitions. They read
(the explicit expression for the momentum canonically conjugate to $a$
is not needed in what follows)
\begin{equation}P_b = -\frac{a^2}{6\lP^2 N} \int \dd^3 x
\gamma^{1/2}\, w_{ij}w^{ij},
\label{pb}
\end{equation}
\begin{equation}
P_{_N} = \frac{ba^2}{6\lP^2 N^2} \int \dd^3 x \gamma^{1/2}\, w_{ij} w^{ij}
\label{pn}
\end{equation}
and
\begin{equation}
\Pi^{ij} = \frac{a^3\gamma^{1/2}\dot{w}^{ij}}{12\lP^2 N}.
\end{equation}
Equations (\ref{pb}) and (\ref{pn}) are constraints, which must be
added to the Hamiltonian via Lagrange multipliers $\alpha$ and
$\beta$. In order to simplify the treatment of the constraints, we
make the following canonical transformations:
\begin{eqnarray}
\tilde{P}_{_N} & = & P_{_N}+\frac{bP_b}{N},\\
\tilde{N} & = & N,\\
\tilde{P}_b & = & N P_b,\\
\tilde{b} & = & \frac{b}{N},
\end{eqnarray}
and redefine $\tilde{\alpha}\equiv\alpha/N$. After making rescalings
as in Eqs.(\ref{red1})~--~(\ref{red2}) and omitting the tilde, the
Hamiltonian reads
\begin{eqnarray}
H & \equiv & N H_0+\alpha\phi_1+\beta P_{_N} \nonumber\\ & = &
N\left\{ P_a b+b^2 a- k a + \frac{P_{_T}}{a^{3(\lambda-1)}} \left(
1-\frac{1}{4} \int \dd^3 x \gamma^{1/2} \, w_{ij} w^{ij}
\right)^{1-\lambda} \right. \cr && + \frac{b^2 a}{12} \int \dd^3 x
\gamma^{1/2} \, w_{ij} w^{ij} \nonumber\\ & & + \left. \int \dd^3 x
\left[ \gamma^{1/2} \left( \frac{a w_{ij|k} w^{ij|k}}{24} + \frac{k a
w_{ij} w^{ij}}{6} \right) + 6\frac{\Pi^{ij}
\Pi_{ij}}{a^{3}\gamma^{1/2}} \right] \right\} \cr & &
+\alpha\left(P_b+\frac{a^2}{6} \int \dd^3 x \gamma^{1/2}\, w_{ij}
w^{ij} \right)+ \beta P_{_N}.
\label{h5}
\end{eqnarray}

Demanding the conservation of these constraints, that is, the
vanishing of their Poisson Brackets with the Hamiltonian (\ref{h5}),
we obtain the following secondary constraints
\begin{equation}
\label{f27}
\phi_{2} \equiv -2ba + \frac{ba}{6} \int \dd^3 x \gamma^{1/2}\, w_{ij}
w^{ij} - P_a + \frac{4}{a}\int \dd^3 x w_{ij} \Pi^{ij} \approx 0,
\end{equation}
and
\begin{equation}
H_0\approx 0.
\end{equation}
Conservation of the secondary constraints $\phi_{2}\approx 0$ then fixes
the value of $\alpha$ through
\begin{equation}
\alpha = -\frac{\{\phi_2,H_0\}}{\{\phi_2,\phi_1\}},
\end{equation}
where $\{,\}$ stands for Poisson brackets.

It is straightforward to check that $P_{_N}$ and $H_0$ are first class
constraints, \ie they have vanishing Poisson Brackets with all other
constraints, while $\phi_1$ and $\phi_2$ are second class constraints,
which is the reason why the unambiguous determination of $\alpha$ is
possible. The way to deal with these second class constraints is to
work in the formalism of the Dirac brackets.  They are defined as
\begin{equation}
\{A,B\}^\mathrm{D} = \{A,B\} - \{A,\phi_i\} (C^{-1})^{ij}\{\phi_j,B\},
\label{pdd}
\end{equation}
where $(C^{-1})^{ij}$ is the inverse of the antisymmetric matrix
$C_{ij}\equiv\{\phi_i,\phi_j\}$. In our case
\begin{equation}
C_{12}=-C_{21}=2a\left( 1 + \frac{5}{12} \int \dd^3 x \gamma^{1/2}\,
w_{ij} w^{ij}\right).
\label{pd}
\end{equation}
By definition, the Dirac brackets of a second class constraint with
any phase space function is zero, so we have
$\{A,\phi_i\}^\mathrm{D}=0$, and one can view the second class
constraints as strong equalities.  Hence, from Eq.~(\ref{f27}) one can
write (always up to second order in the perturbations)
\begin{equation}
b=-\frac{P_a}{2a} + \frac{2}{a^2} \int \dd^3 x w_{ij} \Pi^{ij} -
\frac{P_a}{24 a} \int \dd^3 x \gamma^{1/2} w_{ij} w^{ij},
\label{b}
\end{equation}
and substitute it into the Hamiltonian (\ref{h5}), yielding
\begin{eqnarray}
H & = & N \left\{ - \frac{P_a^2}{4a} - ka
+\frac{P_{_T}}{a^{3(\lambda-1)}} + \frac{1}{12} \left[
\frac{P_a^2}{4a} + 3 (\lambda-1) \frac{P_{_T}}{a^{3(\lambda-1)}} + 2 k
a \right] \int \dd^3 x \gamma^{1/2}\, w_{ij} w^{ij}\right. \nonumber\\
& & + \left. \int \dd^3 x \left( \frac{\gamma^{1/2} a}{24} w_{ij|k}
w^{ij|k} + 6\frac{\Pi^{ij}
\Pi_{ij}}{a^{3}\gamma^{1/2}}\right)\right\}.
\label{h6}
\end{eqnarray}

Substituting Eq.~(\ref{pd}) in Eq.~(\ref{pdd}), we get the following
Dirac Brackets between the dynamical variables (up to second order in
the perturbations):
\begin{eqnarray}
\{a,\Pi^{ij}(x)\}^\mathrm{D}&=&\frac{1}{6} a w^{ij}(x), \label{d1}\\
\{a,w_{ij}(x)\}^\mathrm{D}&=&0,\\
\{a,P_a\}^\mathrm{D}&=&1+\frac{Q}{6},\\ \{a,P_{_T}\}^\mathrm{D}&=&0,\\
\{a,T\}^\mathrm{D}&=&0,\\ \{P_a,\Pi^{ij}(x)\}^\mathrm{D}&=&\frac{1}{6}
\gamma^{1/2}P_a w^{ij}(x),\\ \{P_a,w_{ij}(x)\}^\mathrm{D}&=&0,\\
\{P_a,P_{_T}\}^\mathrm{D}&=&0,\\ \{P_a,T\}^\mathrm{D}&=&0,\\
\{T,w_{ij}\}^\mathrm{D}&=&0,\\ \{T,\Pi^{ij}(x)\}^\mathrm{D}&=&0,\\
\{T,P_{_T}\}^\mathrm{D}&=&1,\\
\{w_{kl}(x),\Pi^{ij}(x')\}^\mathrm{D}&=&\delta
^{ij}_{kl}\delta^3(x-x')- \frac{2}{3}w_{kl}(x)w^{ij}(x'),
\label{d13}
\end{eqnarray}
where we have used the definition (\ref{defQ}) for the quantity $Q$.

The time evolution of any phase space function is now given by
\begin{equation}
\label{dA}
\dot{A}=\{A,H\}^\mathrm{D}.
\end{equation}
Hence we will define phase space functions which have canonical Dirac
brackets among themselves, namely 
\begin{equation}
\{T_{({\rm c})},P_{({\rm c})_T}\}^\mathrm{D}=1, \qquad \qquad
\{a_{({\rm c})},P_{({\rm c})a}\}^\mathrm{D}=1, 
\end{equation}
and
\begin{equation}
\{w_{({\rm c}) kl}(x),\Pi_{({\rm c})}^{ij}(x')\}^\mathrm{D}= \delta
^{ij}_{kl}\delta^3(x-x'),
\end{equation}
the others vanishing. The different new canonical functions are
\begin{eqnarray}
a_{({\rm c})} & = & a\left(1-\frac{Q}{12}\right),\\
&&\nonumber\\
P_{({\rm c})a}& = & P_a\left(1-\frac{Q}{12}\right),\\
w_{({\rm c})ij}& = & w_{ij}\left(1+\frac{Q}{3}\right).
\end{eqnarray}
If we now express the Hamiltonian (\ref{h6}) in terms of these
canonical functions, we obtain, omitting the index (c),
\begin{eqnarray}
H = N H_0 &=& N\left[-\frac{P_a^2}{4a} - k a +
\frac{P_{_T}}{a^{3(\lambda - 1)}} + \int \dd^3 x \left( 6
\frac{\Pi^{ij} \Pi_{ij}}{\gamma^{1/2}a^{3}} + \frac{1}{24}
\gamma^{1/2} a w_{ij|k} w^{ij|k}\right.\right.\cr & & +\left.\left.
\frac{1}{12} \gamma^{1/2} k w_{ij} w^{ij}a\right)\right],
\label{hc1}
\end{eqnarray}
which is exactly Hamiltonian (\ref{h2}) of the preceding section, also
obtained in the previous subsection through the canonical
transformations (\ref{tc20}).  Note that at the level of the equations
of motion, which are first order in the perturbations, there is no
difference between variables with or without the index (c). With the
time evolution law (\ref{dA}), and remembering that (\ref{hc1}) is
expressed in terms of phase space functions with canonical Dirac
brackets relations, it is obvious that (\ref{hc1}) generates the
classical GR equations for this system.  Hence, without using any
background equations of motion, we were able to obtain the simpler
Hamiltonian (\ref{hc1}), which is equivalent to (\ref{h2}), from the
more complicated form (\ref{h1}), as we did in the last subsection
using canonical transformations directly in (\ref{h1}).

\subsection{The functional Schr\"odinger equation}

As we are here also quantizing the background, the quantization
procedure is now to impose ${\hat{H}}_0\Psi(a,w_{ij})=0$, with
${\hat{H}}_0$ being the operator version of Eq.~(\ref{hc1}), where the
operator algebra comes from $[,]=i\hbar\{,\}$ in the Hamiltonian point
of view, or $[,]=i\hbar\{,{\}}^\mathrm{D}$ in the Lagrangian case,
which of course turn to be the same.  Evidently, ${\hat{H}}_0$
stemming from Eq.~(\ref{hc1}) is much simpler than that coming from
Eq.~(\ref{h1}). We will from now on focus our attention on
Eq.~(\ref{hc1}).

The Wheeler-DeWitt equation in this case, with the particular factor
ordering of Eq.~(\ref{h200}) for ${\hat{H}}_0$, reads
\begin{eqnarray}
\left\{ \frac{1}{4a} \frac{\partial^2}{\partial a^2} \right. && - k a
- i\frac{1}{a^{3(\lambda-1)}} \frac{\partial}{\partial T} + \int \dd^3
x \left[ -6 \frac{1}{a^3\gamma^{1/2}} \frac{\delta^2}{\delta
w_{ij}\delta w^{ij}} \right. \cr && \left. \left. + a
\left(\gamma^{1/2}\frac{w_{ij|k} w^{ij|k}}{24} + k \frac{w_{ij}
w^{ij}}{12}\right)\right]\right\}\Psi=0.
\label{es15} \end{eqnarray}

When the matter fields are described by a perfect fluid,
there appears a first order derivative with respect to the
variable describing the fluid, which enforces us to interpret
it as the time variable on which the wave functional evolves.
This is a particular feature of the fluid description, not present
when the matter fields are described by, e.g., a scalar field.
Choosing $T$ as the time variable is equivalent to impose
de time gauge $N=a^{3(\lambda-1)}$. Eq.~(\ref{es15}) then reads
\begin{eqnarray}
i\frac{\partial\Psi}{\partial T} & = & \hat{H}_\mathrm{red} \Psi
\nonumber \\ &:=&\left\{ \frac{a^{3\lambda-4}}{4}
\frac{\partial^2}{\partial a^{2}} - k a^{3\lambda-2}+ \int \dd^3 x
\left[ - 6 \frac{a^{3(\lambda-2)}}{\gamma^{1/2}}
\frac{\delta^2}{\delta w_{ij} \delta w^{ij}} \right.\right.
\cr & & \left. \left. + a^{3\lambda-2}
\left(\gamma^{1/2} \frac{w_{ij|k} w^{ij|k}}{24} + k\frac{w_{ij}
w^{ij}}{12}\right)\right]\right\}\Psi.
\cr & & 
\label{es2}
\end{eqnarray}

\section{Further Developments}
\label{sec:further}

In this section we begin by further simplifying the functional
Schr\"odinger equation (\ref{es2}) with the use of the Bohm-de Broglie
interpretation of quantum mechanics, and then we present, for the sake
of completeness, the scalar field case.

\subsection{Further simplification of the wave functional equation
using the Bohm-de Broglie interpretation}

Note that Eq.~(\ref{es2}), although simpler than the one that would
have been obtained through a quantization procedure using Hamiltonian
(\ref{h1}), has not yet its gravitational wave part in the simplest
form (\ref{schro1}). In fact, Eq.~(\ref{schro1}) was obtained in the
previous section through the time dependent quantum canonical
transformation generated by Eq.~(\ref{qct0}), which transformed the
operator version of Eq.~(\ref{hgw2}) into the operator version of
Eq.~(\ref{hgw3}). This was done under the assumption that $a(T)$ was a
given function of the time parameter. However, in a non ontological
interpretation of quantum mechanics, the function $a(T)$ does not make
sense when $a$ is also quantized since trajectories do not exist in
this case. Hence, Eq.~(\ref{es2}) is the ultimate form that can be
achieved in such interpretations.

On the other hand, if one uses an ontological interpretation of
quantum mechanics like the one suggested by de Broglie and
Bohm~\cite{onto}, and makes the separation ansatz for the wave
functional $$\Psi[a,w_{ij},T]=\varphi(a,T)\psi[a,w_{ij},T],$$ with
$$\psi[a,w_{ij},T]=\psi_1[w_{ij},T]\int \dd a \varphi^{-2}(a,T)
+\psi_2[w_{ij},T],$$ then Eq.~(\ref{es2}) can be split into two,
yielding
\begin{equation}
i\frac{\partial\varphi}{\partial T}
=\frac{a^{3\lambda-4}}{4} \frac{\partial^2\varphi}{\partial a^2}-
ka^{3\lambda-2}\varphi,
\label{es20}
\end{equation}
and
\begin{equation}
i\frac{\partial\psi}{\partial T}= \int \dd^3 x
\left[-6\frac{a^{3(\lambda-2)}}{\gamma^{1/2}} \frac{\delta^2}{\delta
w_{ij} \delta w^{ij}} + a^{3\lambda-2} \left(\gamma^{1/2}
\frac{w_{ij|k} w^{ij|k}}{24} + k\frac{w_{ij}
w^{ij}}{12}\right)\right]\psi.
\label{es22}
\end{equation}

Using the Bohm interpretation, Eq.~(\ref{es20}) can now be solved as
in Ref.~\cite{pinto,pinto2,fabris}, yielding a Bohmian quantum
trajectory $a(T)$, which in turn can be used in
Eq.~(\ref{es22}). Indeed, since one can view $a(T)$ as a function of
$T$, it is possible to apply the canonical transformations (\ref{ct2})
generated by Eq.~(\ref{f2}) in the same way as in the previous
section. With these transformations, Eq.~(\ref{es22}) reads
\begin{equation}
i\frac{\partial\psi}{\partial T}=
\int \dd^3 x \left[-\frac{a^{3\lambda-4}}{2\gamma^{1/2}}
\frac{\delta^2}{\delta \mu_{ij}\delta \mu^{ij}} + a^{3\lambda-4}
\left(\gamma^{1/2} \frac{\mu_{ij|k} \mu^{ij|k}}{2} +
k \mu_{ij}\mu^{ij} - \frac{\ddot{a}}{2a} \mu_{ij}
\mu^{ij}\right)\right] \psi,
\label{es27}
\end{equation}
Through the redefinition of time $a^{3\lambda-4}\dd T=\dd\eta$, we
recover Eq.~(\ref{schro1}), namely
\begin{eqnarray}
i\frac{\partial\psi(\mu_{ij},\eta)}{\partial \eta}&=&
\int \dd^3 x \left\{-\frac{1}{2\gamma^{1/2}}
\frac{\delta^2}{\delta \mu_{ij}\delta \mu^{ij}} +
\gamma^{1/2}\left[\frac{1}{2}\mu_{ij|k} \mu^{ij|k} + \left( k-
\frac{\ddot{a}}{2a}\right) \mu_{ij} \mu^{ij}\right] \right\} \cr \cr
&\times& \psi(\mu_{ij},\eta).
\end{eqnarray}

This is the most simple form of the Schr\"odinger equation which
governs tensor perturbations for fluid matter source.  In this way, we
can proceed with the usual analysis, with the quantum Bohmian solution
$a(\eta)$ coming from Eq.~(\ref{es20}) acting as the new pump field.

If the fluid is radiation, which is one of the best phenomenological
approach for the primordial Universe, then $\lambda=\frac{4}{3}$, and
there is no factor ordering ambiguity neither in the operator version
of Eq.~(\ref{hc1}) (in this case conformal time is obtained through
setting $N=a$) nor in Eq.~(\ref{es20}).  Furthermore, in a restricted
Hilbert space, the operator read in the right-hand-side of
Eq.~(\ref{es20}) can be made self adjoint~\cite{lemos}.

\subsection{The scalar field case}\label{sec:scalarfield}

In order to be complete, we end this section by a rapid examination of
the specific case of a scalar field to exhibit specifically the
differences with the fluid case.

In the case of a minimally coupled scalar field, the scalar field
action which must be added to the gravitational action
(\ref{graction}) reads
\begin{eqnarray}
S_{\phi}&=&\int \dd^4 x \sqrt{-g}\left[ \frac{1}{2} \partial_{\mu} \phi
\partial^{\mu} \phi - V (\phi)\right] \cr &=&\int \dd^4 x \gamma^{1/2}
a^3\left\{ \frac{{\dot{\phi}}^2}{2N} - N V(\phi) - \frac{1}{4} w_{ij}
w^{ij} \left[\frac{{\dot{\phi}}^2}{2N} - N V(\phi)\right]
\right\}, \label{fiaction}
\end{eqnarray}
where we have used the form (\ref{adm2}) of the perturbed ADM metric
and expanded the action to second order in the perturbations as before.

Calculating the Hamiltonian in the standard way, and performing
redefinitions as in Eqs.~(\ref{red2}), we obtain
\begin{eqnarray}H & \equiv & N H_0\nonumber \\
& = & N\left\{ - \frac{P_a^2}{4a} - k a + \frac{\Pi_{\phi}^2}{2a^3} +
a^3 V(\phi)\left( 1-\frac{1}{4}\int \dd^3 x \gamma^{1/2}\, w_{ij}
w^{ij}\right) \right. \cr && + \frac{5P_a^2}{48a} \int \dd^3 x
\gamma^{1/2}\, w_{ij} w^{ij}\nonumber\\&& +\int \dd^3 x \left. \left[
\frac{6\tilde{\Pi}_{ij}\tilde{\Pi}^{ij}}{a^{3}\gamma^{1/2}} +
2\frac{P_a w_{ij}\tilde{\Pi}^{ij}}{a^{2}} + \gamma^{1/2} a \left(
\frac{w^{ij|k} w_{ij|k}}{24} + \frac{k}{6}w_{ij} w^{ij},
\right)\right]\right\},
\label{hfi}
\end{eqnarray}
where $\Pi_{\phi}$ is the momentum canonically conjugated to $\phi$;
this is exactly the Hamiltonian obtained in Ref.~\cite{halli} for
tensor perturbations.

After performing the canonical transformations of subsection
\ref{sec:Hclq}, we obtain finally
\begin{eqnarray}
H&=&N\biggl[-\frac{P_a^2}{4a}  - k a + \frac{\Pi_\phi^2}{2a^3} +
a^3 V(\phi) \cr &&+ \left. \int \dd^3 x \left( 6
\frac{\Pi^{ij}\Pi_{ij}}{\gamma^{1/2}a^3} +
\frac{1}{24}\gamma^{1/2}aw_{ij|k}w^{ij|k}+
\frac{1}{12}\gamma^{1/2}kw_{ij}w^{ij}a\right)\right],
\label{hfi2}
\end{eqnarray}
which is, as before, a great simplification over Eq.~(\ref{hfi}).
Note, however, that, in the case of the scalar field, there is no
linear momentum which can be obviously chosen as a time derivative as
in the perfect fluid case [see Eq.~(\ref{h2}) for comparison].  To
obtain equations like (\ref{es20}) and (\ref{es22}), and then
(\ref{es27}), one must make a choice of time, which is rather
arbitrary~\cite{isham}, and problematic~\cite{kuchar} in this
situation. Anyway, at the semiclassical level it is certain that
Eq.~(\ref{es27}) can be obtained adopting procedures similar to what
was done in Ref.~\cite{halli} for the quantum equation coming from
Hamiltonian (\ref{hfi2}).

\section{Conclusion} \label{sec:conc}

Considering the case for which the background is quantized, we have
obtained the form of the quantum equations governing tensor
perturbations. The result can be made as simple as when the background
satisfies Einstein classical field equations. This we achieved in two
successive steps.

The first step is general and the simplification is obtained through
either classical and quantum canonical transformations in the
Hamiltonian point of view, or through the elimination of a total time
derivative and the subsequent Ostrogradski treatment of second time
derivatives which appear in the lagrangian formalism.

When the matter source is described by a perfect fluid and a notion of
time appears naturally, the second step demands, in order to be
implemented, using the Bohm-de Broglie interpretation of quantum
mechanics. In this framework, a real quantum trajectory $a(t)$ in
minisuperspace does exist, and a further quantum canonical
transformation can be made in order to put the quantum equations
exactly in the same form that appears in most previous works dealing
with the evolution of cosmological tensor perturbations in classical
backgrounds. The only difference is that the time dependent effective
mass is no longer calculated in terms of the classical background
solution for $a(t)$, but in terms of the quantum Bohmian solution
$a(t)$, which have been obtained in all fluid
cases~\cite{pinto,fabris}. Hence, in the Bohmian approach, the quantum
background effects turn out to be very simple to calculate, much
easier than in other, non ontological, interpretations of quantum
mechanics.

It is interesting to ask the question as to whether the Bohmian
approach, which turns out to be also the simplest for this situation,
will provide solutions for the observable predictions that differ from
the other interpretations of quantum mechanics applicable to quantum
cosmology~\cite{MW}. It could be conjectured that self-consistency of
quantum principles would imply that the outcome of the calculation
should eventually not depend of the interpretation, so that using the
Bohm trajectories would then be reducible merely to a clever
computational trick. However, unless the actual calculation is
performed, the opposite can also be conjectured, namely that there
could exist different predictions in the quantum cosmological
framework that could allow, in principle and thus possibly
observationnally, to discriminate between various alternatives.  The
case of matter being described by a scalar field, as was briefly
discussed here, needs further elaboration: the problem here lies in
the absence of a natural time. Whatever the answer to the
abovementioned problems, it is clear that the framework of Bohm
interpretation seems adequate in order to go further.

Whether similar calculations and simplification can be implemented in
the scalar \cite{dl} and vector perturbation cases is rather doubtful
and yet under study~\cite{future}. We guess that the simplification
obtained here, completely independent of the background equations, is
a feature of tensor perturbations, and tensor only, therefore most
presumably not to be generalized to scalar and vector
perturbations. In fact, we believe this is related to the fact that
only tensor perturbations do not have any linear perturbation terms
present in the action, as such terms happen to vanish
identically. This is not the case, e.g., of the action of scalar
perturbations, which does contain such linear terms, with coefficients
that vanish provided the background functions satisfy the background
equations of motion, as it should.  Also, the form of the equations of
motion governing the dynamics of tensor perturbations is completely
independent of the matter source of the gravitational field, contrary,
e.g., to the scalar perturbations case.

\section{Acknowledgements}

Two of us (EP and NPN) would like to thank CNPq of Brazil for
financial support. PP should like to acknowledge the Centro Brasileiro
de Pesquisas F\'{\i}sicas (CBPF), for warm hospitality during the time
this work was initiated. NPN would similarly like to thank the Groupe
de Relativit\'e et de Cosmology (${\cal G}\setR\varepsilon\setC{\cal
O}$) at the Institut d'Astrophysique de Paris, where this work was
completed. We would like to thank G.~Esposito-Far\`ese and J.~Martin
for various enlightening discussions. We also would like to thank
CAPES (Brazil) and COFECUB (France) for partial financial support.

\end{document}